\documentclass[prc,twocolumn,10pt]{revtex4}
\usepackage{epsf,epsfig,latexsym}
\usepackage{amsmath}
\usepackage{amssymb}
\usepackage{bm}
\usepackage{graphicx}
\usepackage{dcolumn}
\usepackage{multirow}
\usepackage{units}

\DeclareRobustCommand{\dgr}[1][]{
  \unit[#1]{\ifmmode{}^\circ\else${}^\circ$\fi}}

\def\ssNN#1{\sqrt{s_{NN}} \ifx|#1|\else=\unit[#1]{GeV}\fi}
\DeclareRobustCommand{\snn}[1]{\ifmmode\ssNN{#1}\else$\ssNN{#1}$\fi}
\DeclareRobustCommand{\valn}[3]{\ifmmode #1\,\pm\,_{#2\text{ (stat)}}^{#3\text{ (syst)}}\else$#1\,\pm\,_{#2\text{ (stat)}}^{#3\text{ (syst)}}$\fi}
\DeclareRobustCommand{\val}[3]{\ifmmode #1\,\pm\,_{#2}^{#3}\else$#1\,\pm\,_{#2}^{#3}$\fi}

\DeclareRobustCommand{\mpt}{\ifmmode\left<p_T\right>\else$\left<p_T\right>$\fi}
\DeclareRobustCommand{\pT}{\ifmmode p_T\else$p_T$\fi}
\DeclareRobustCommand{\mTm}{\ifmmode m_T-m\else$m_T-m$\fi}

\newcommand{\bnl}           {$\rm^{1}$}
\newcommand{\ires}          {$\rm^{2}$}
\newcommand{\kraknuc}       {$\rm^{3}$}
\newcommand{\krakow}        {$\rm^{4}$}
\newcommand{\baltimore}     {$\rm^{5}$}
\newcommand{\newyork}       {$\rm^{6}$}
\newcommand{\nbi}           {$\rm^{7}$}
\newcommand{\texas}         {$\rm^{8}$}
\newcommand{\bergen}        {$\rm^{9}$}
\newcommand{\bucharest}     {$\rm^{10}$}
\newcommand{\kansas}        {$\rm^{11}$}
\newcommand{\oslo}          {$\rm^{12}$}

\begin{document}
\bibliographystyle{apsrev}

\title{Charged meson rapidity distributions in central Au+Au collisions at
  \snn{200}}

\author{
  I.~G.~Bearden\nbi,
  D.~Beavis\bnl,
  C.~Besliu\bucharest,
  B.~Budick\newyork,
  H.~B{\o}ggild\nbi,
  C.~Chasman\bnl,
  C.~H.~Christensen\nbi,
  P.~Christiansen\nbi,
  J.~Cibor\kraknuc,
  R.~Debbe\bnl,
  E. Enger\oslo,
  J.~J.~Gaardh{\o}je\nbi,
  M.~Germinario\nbi,
  K.~Hagel\texas,
  O.~Hansen\nbi,
  A.~Holm\nbi,
  A.~K.~Holme\oslo,
  H.~Ito\kansas,
  A.~Jipa\bucharest,
  F.~Jundt\ires,
  J.~I.~J{\o}rdre\bergen,
  C.~E.~J{\o}rgensen\nbi,
  R.~Karabowicz\krakow,
  E.~J.~Kim\kansas,
  T.~Kozik\krakow,
  T.~M.~Larsen\oslo,
  J.~H.~Lee\bnl,
  Y.~K.~Lee\baltimore,
  G.~L{\o}vh{\o}iden\oslo,
  Z.~Majka\krakow,
  A.~Makeev\texas,
  M.~Mikelsen\oslo,
  M.~Murray\kansas,
  J.~Natowitz\texas,
  B.~S.~Nielsen\nbi,
  J.~Norris\kansas,
  K.~Olchanski\bnl,
  D.~Ouerdane\nbi,
  R.~P\l aneta\krakow,
  F.~Rami\ires,
  C.~Ristea\bucharest,
  D.~R{\"o}hrich\bergen,
  B.~H.~Samset\oslo,
  D.~Sandberg\nbi,
  S.~J.~Sanders\kansas,
  R.~A.~Sheetz\bnl,
  P.~Staszel\nbi,
  T.~S.~Tveter\oslo,
  F.~Videb{\ae}k\bnl,
  R.~Wada\texas,
  Z.~Yin\bergen, and
  I.~S.~Zgura\bucharest\\
  (BRAHMS Collaboration )\\[1ex]
  \bnl~Brookhaven National Laboratory, Upton,New York 11973,\\
  \ires~Institut de Recherches Subatomiques and Universit{\'e} Louis
  Pasteur, Strasbourg, France,\\
  \kraknuc~Institute of Nuclear Physics, Krakow, Poland,\\
  \krakow~Jagiellonian University, Krakow, Poland,\\
  \baltimore~Johns Hopkins University, Baltimore, Maryland 21218,\\
  \newyork~New York University, New York, New York 10003,\\
  \nbi~Niels Bohr Institute, University of Copenhagen, Denmark,\\
  \texas~Texas A$\&$M University, College Station,Texas 77843,\\
  \bergen~University of Bergen, Department of Physics, Bergen,Norway,\\
  \bucharest~University of Bucharest,Romania,\\
  \kansas~University of Kansas, Lawrence, Kansas 66049,\\
  \oslo~University of Oslo, Department of Physics, Oslo,Norway}
\noaffiliation

\begin{abstract}
  We have measured rapidity densities $dN/dy$ of $\pi^{\pm}$ and
  $K^{\pm}$ over a broad rapidity range ($-0.1 < y < 3.5$) for central
  Au+Au collisions at \snn{200}. These data have significant
  implications for the chemistry and dynamics of the dense system that
  is initially created in the collisions.  The full phase--space
  yields are $1660\pm 15\pm 133$ ($\pi^+$), $1683\pm 16\pm 135$
  ($\pi^-$), $286\pm 5\pm 23$ ($K^+$) and $242\pm 4\pm 19$ ($K^-$).
  The systematics of the strange to non--strange meson ratios are
  found to track the variation of the baryo--chemical potential with
  rapidity and energy. Landau--Carruthers hydrodynamic is found to
  describe the bulk transport of the pions in the longitudinal
  direction.
\end{abstract}
\maketitle


In ultra--relativistic heavy ion collisions at RHIC energies, charged
pions and kaons are produced copiously. The yields of these light
mesons are indicators of the entropy and strangeness created in the
reactions, sensitive observables to the possible existence of an early
color deconfined phase, the so--called quark gluon plasma. In such
collisions, the large number of produced particles and their
subsequent reinteractions, either at the partonic or hadronic level,
motivate the application of concepts of gas or fluid dynamics in
their interpretation. Hydrodynamical properties of the expanding
matter created in heavy ion reactions have been discussed by
Landau~\cite{landau} (full stopping) and Bjorken~\cite{bjorken}
(transparency), in theoretical pictures using different initial
conditions. In both scenarios, thermal equilibrium is quickly achieved
and the subsequent isentropic expansion is governed by hydrodynamics.
The relative abundances and kinematic properties of particles provide
an important tool for testing whether equilibrium occurs in the course
of the collision.  In discussing the source characteristics, it is
important to measure most of the produced particles in order not to
violate conservation laws (e.g.  strangeness and charge conservation).

In this letter, we report on the first measurements at RHIC energies
of transverse momentum ($\pT$) spectra of $\pi^\pm$ and $K^\pm$ over
the rapidity range $-0.1 < y < 3.5$ for the 5\% most central Au+Au
collisions at \snn{200}.  The spectra are integrated to obtain yields
as a function of rapidity ($dN/dy$), giving full phase--space ($4\pi$)
yields. At RHIC energies, a low net--baryon density is observed at
mid--rapidity~\cite{peter}, so mesons may be predominantly produced
from the decay of the strong color field created initially.  At
forward rapidities, where primordial baryons are more
abundant~\cite{ratio200}, other production mechanisms, for example
associated strangeness production, play a larger role. Therefore, the
observed rapidity distributions provide a sensitive test of models
describing the space--time evolution of the reaction, such as Landau
and Bjorken models~\cite{landau,bjorken}. In addition, integrated
yields are a key input to statistical models of particle
production~\cite{becat,braun}.\\

BRAHMS consists of two hadron spectrometers, a mid--rapidity arm (MRS)
and a forward rapidity arm (FS), as well as a set of detectors for
global event characterization~\cite{nim}. Collision centrality is
determined from charged particle multiplicities, measured by
scintillator tile and silicon multiplicity arrays located around the
nominal interaction point. The interaction vertex is measured with a
resolution of \unit[0.6]{cm} by arrays of \v{C}erenkov counters
positioned on either side of the nominal vertex. Particle
identification (PID) for momenta below \unit[2]{GeV/c} is performed
via time--of--flight (TOF) in the MRS. In the FS, TOF capabilities
allow $\pi$--$K$ separation up to $p = \unit[4.5]{GeV/c}$, and is
further extended up to $\unit[20]{GeV/c}$ using a ring imaging
\v{C}erenkov detector. Further details can be found
in~\cite{nim,myself}.

\begin{figure}[htb]
  \centering
  \includegraphics[width=\columnwidth]{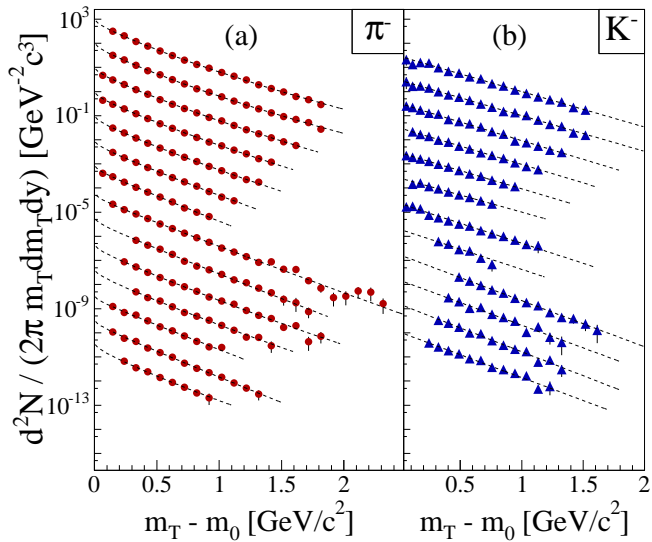}
  \caption{Invariant transverse mass $m_T - m_0$ spectra of $\pi^-$
    (a) and $K^-$ (b) from $y\sim 0$ (top) to $y\sim 3.5$ (bottom).
    Dashed lines are fits to the data, namely a power law in $p_T$ for
    pions and an exponential in $m_T - m_0$ for kaons. Errors are
    statistical. Spectra have been rescaled by powers of 10 for
    clarity.}
  \label{spec}
\end{figure}

Figure~\ref{spec} shows transverse mass $m_T-m_0$ spectra ($m_T =
\sqrt{p_T^2 + m_0^2}$) for $\pi^-$ and $K^-$.
Particle spectra were obtained by combining data from several
spectrometer settings (magnetic field and angle), each of which covers
a small region of the phase--space $(y,p_T)$. The data have been
corrected for the limited acceptance of the spectrometers using a
Monte-Carlo calculation simulating the geometry and tracking of the
BRAHMS detector system. Detector efficiency, multiple scattering and
in--flight decay corrections have been estimated using the same
technique. Hyperon ($\Lambda$) and neutral kaon $K_{0s}$ decays may
have contaminated the pion sample. For $K_{0s}$, it is assumed that its
yields amount to the average between $K^+$ and $K^-$ at each rapidity
interval. For $\Lambda$ yields, since only mid--rapidity data are
available, we used the same assumptions as in~\cite{peter}, namely
$\Lambda/p = \bar{\Lambda}/\bar{p} = 0.9$ in the phase--space covered
in this analysis. The fraction of pions originating from $\Lambda$ and
$K_{0s}$ decays was estimated with a GEANT simulation where realistic
particle distributions (following an exponential in $m_T$) were
generated for several spectrometer settings. Particles were tracked
through the spectrometers and produced pions were accepted according
to the same data cuts applied to the experimental data. It has been
found a $K_{0s}$ ($\Lambda$) contamination of 4\% ($\lesssim$~1\%) in
the MRS and 6\% ($\lesssim$~1\%) in the forward spectrometer. In the
following, the pion yields are corrected unless stated otherwise.

The pion spectra are well described at all rapidities by a power law
in $p_T$, $A (1 + p_T/p_0)^{-n}$. For kaons, an exponential in $m_T -
m_0$, $A \exp\left(\frac{m_T - m_0}{T}\right)$, has been used. The
invariant yields $dN/dy$ were calculated by integrating the fit
functions over the full $p_T$ or $m_T$ range. The two main sources of
systematic error on $dN/dy$ and \mpt{} are the extrapolation in the
low $p_T$ range outside the acceptance, and the normalization of the
spectra. Other fit functions were used in order to estimate the error
on the extrapolation. In the FS, due to a smaller acceptance coverage
at low $p_T$, the error is systematically larger than in the MRS. In
total, the systematic error amounts to $\sim$ 10\% in the range $-0.1
< y < 1.4$ (MRS) and $\sim$ 15\% for $y > 2$ (FS).  Mid--rapidity
yields recently reported by the STAR~\cite{star200} and PHENIX
experiments~\cite{phenix200} are within $1\,\sigma_{syst}$ of these
results.

\begin{figure}[h]
  \centering
  \includegraphics[width=\columnwidth]{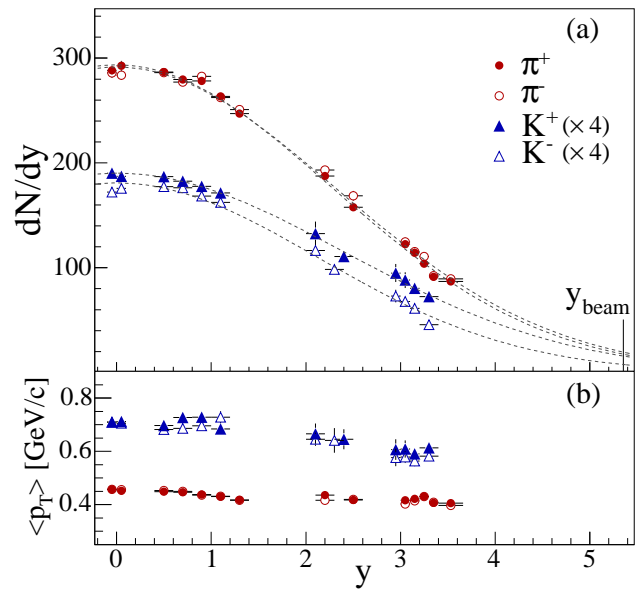}
  \caption{Pion and kaon rapidity densities (a) and their mean
    transverse momentum \mpt{} (b) as a function of rapidity.  Errors
    are statistical. The kaon yields were multiplied by 4 for clarity.
    The dashed lines in (a) are Gaussian fits to the $dN/dy$
    distributions (see text).}
  \label{fig:dndy}
\end{figure}

Rapidity densities and mean transverse momenta \mpt{} extracted from
the fits are shown in Fig.~\ref{fig:dndy}.
Panel (a) shows the pion and kaon yields. $\pi^+$ and $\pi^-$ are
found in nearly equal amounts within the rapidity range covered, while
an excess of $K^+$ over $K^-$ is observed to increase with
rapidity~\cite{datatab}.  Figure~\ref{fig:dndy}(b) shows the rapidity
dependence of \mpt{}.  There is no significant difference between
positive and negative particles of a given mass. For pions, $\mpt{} =
\unit[0.45 \pm 0.05]{GeV/c}\,$(stat~+~syst) at $y = 0$ and decreases
little to $\unit[0.40 \pm 0.06]{GeV/c}$ at $y = 3.5$, while for kaons,
\mpt{} drops from $\unit[0.71 \pm 0.07]{GeV/c}$ at $y = 0$ to
$\unit[0.59 \pm 0.09]{GeV/c}$ at $y = 3.3$ (see~\cite{datatab}).

In order to extract full phase space densities for $\pi^\pm$ and
$K^\pm$ we have investigated several fit functions: a single Gaussian
centered at $y=0$ (G1), a sum of two Gaussians (G2) or Woods-Saxon
(WS) distributions placed symmetrically around $y=0$.  Our data do not
distinguish among these functions, all give a $\chi^2$ per d.o.f. of
$\sim 1$, and have the same total integral to within 2\%
(cf. Tab.~\ref{tab:4pimult}).
\begin{table}[htb]
  {\small
    \begin{tabular}{c@{\hspace{0.3cm}}c@{\hspace{0.3cm}}c@{\hspace{0.3cm}}c@{\hspace{0.3cm}}c}\hline\hline
           & $\pi^+$       & $\pi^-$       & $K^+$       & $K^-$\\\hline
        WS & 1677 $\pm$ 17 & 1695 $\pm$ 15 & 293 $\pm$ 6 & 243 $\pm$ 2 \\
        G1 & 1660 $\pm$ 15 & 1683 $\pm$ 16 & 286 $\pm$ 5 & 242 $\pm$ 4 \\
        G2 & 1640 $\pm$ 16 & 1655 $\pm$ 15 & 285 $\pm$ 5 & 239 $\pm$ 3 \\\hline\hline
    \end{tabular}
  }
  \caption{Full phase--space yields of $\pi^\pm$ and $K^\pm$
    extracted from fits to the $dN/dy$ distributions (see
    text). Errors are statistical, systematic errors are of the order
    of 8\%.}
  \label{tab:4pimult}
\end{table}


\begin{figure*}[htb]
  \centering
  \includegraphics[width=0.9\textwidth]{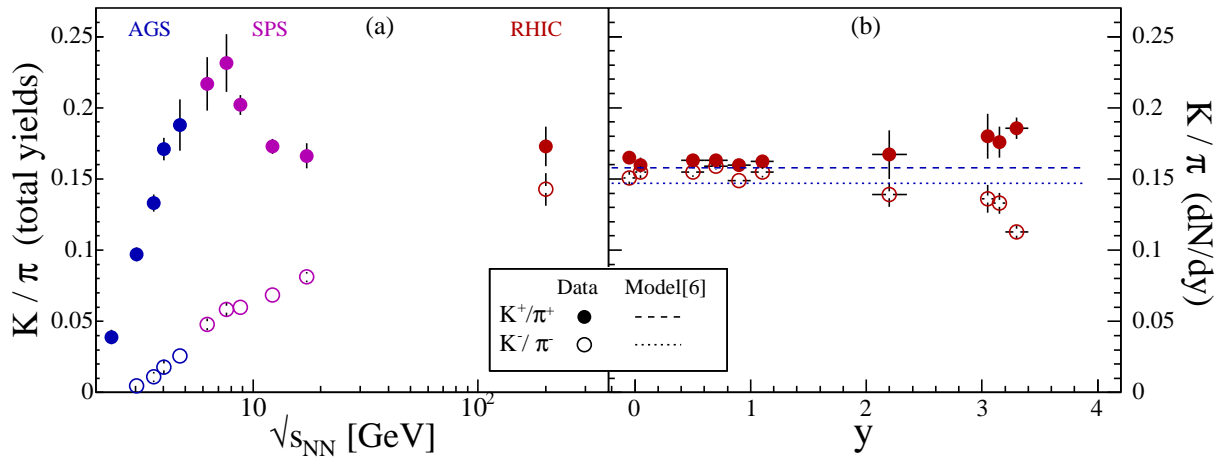}
  \caption{Full phase--space $K/\pi$ ratios as a function of \snn{}
    (a) and rapidity systematics at \snn{200} (b). The dashed and
    dotted lines in (b) are predictions of the statistical
    model~\cite{braun}. Errors are statistical and systematics in (a),
    only statistical in (b). AGS data are from~\cite{kpiAGS1,kpiAGS2},
    SPS data from~\cite{newna49,newna49b,kpiSPS1}. Data points at
    \snn{6.3} and \unit[7.6]{GeV}~\cite{newna49,newna49b} are preliminary.}
  \label{fig:kopi}
\end{figure*}
Figure~\ref{fig:kopi}(a) shows ratios of the strange to non--strange
full phase--space meson yields ($K/\pi$) as a function of \snn{}.  The
ratio $K^+/\pi^+$ shows a fast increase from low AGS to low SPS
energies (\snn{8}), followed by a decrease with increasing energy. At
\snn{200}, we find a value of $0.173 \pm 0.003\pm 0.014$, consistent
with the ratio at the highest SPS energy. In contrast, $K^-/\pi^-$
increases monotonically but remains below $K^+/\pi^+$.  At \snn{200},
it reaches a value of $0.143 \pm 0.003\pm 0.011$, a value close to the
positive ratio. It is interesting to compare the energy systematic
with the rapidity dependence at \snn{200} (Fig.~\ref{fig:kopi}(b)). A
fit to a straight line in the range $0 < y < 1.3$ gives $K^+/\pi^+ =
0.162\pm 0.001$ (stat) and $K^-/\pi^- = 0.154\pm 0.001$. The dashed
and dotted lines are predictions of the hadron gas statistical
model~\cite{braun}, where a chemical freeze--out temperature $T$ of
\unit[177]{MeV} and baryo--chemical potential $\mu_B$ of
\unit[29]{MeV} were used (the authors used a fit to particle ratios at
\snn{130} in the rapidity range $|y|<0.5$ for extrapolating
mid--rapidity ratios at the top RHIC energy). The agreement with the
data is consistent within the systematic errors~\footnote{It should be
  noted that the $\Lambda$ and $K_{0s}$ feed down correction applied
  to the data is significantly different from the amount of feed down
  assumed in the model calculations~\cite{braun}. When not corrected
  for particle feed down, the data show excellent agreement with the
  model prediction.}.  However, the data deviate from the model
prediction at $y > 2$, where an increasing excess of $K^+$ over $K^-$
is observed. This may be due to an increase of the net--baryon
densities~\cite{peter,ratio200}. A baryon rich environment is
favorable for associated strangeness production, e.g.  $p+p\rightarrow
p+K^+ +\Lambda$, a production channel forbidden to $K^-$.  In the
context of the statistical model, this translates into an increase of
the baryo--chemical potential $\mu_B$, as already reported
in~\cite{ratio200}, where a calculation by Becattini {\it et
  al.}~\cite{becat} of $K^+/K^-$ vs $\bar{p}/p$ at constant $T$ (and
varying $\mu_B$) agrees well with the rapidity dependence of the
experimental ratios. \\

\begin{figure}[htb]
  \centering
  \includegraphics[width=\columnwidth]{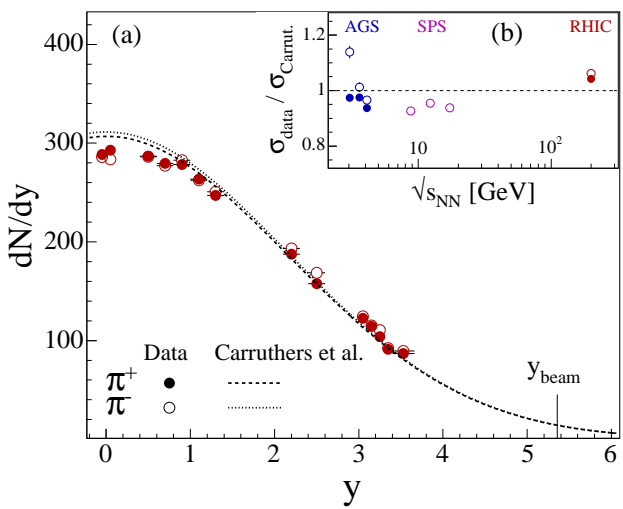}
  \caption{Comparison $dN/dy(\pi)$ and Landau's prediction at
    \snn{200} (a) and ratio $\sigma_{N(\pi)}/\sigma_{\text{Carrut.}}$ as a
    function of $\snn{}$ (b). Errors are statistical.}
  \label{fig:landau}
\end{figure}

The observed rapidity distributions of pions exhibit a nearly Gaussian
shape.  Widths are found to be $\sigma_{\pi^+} = 2.25 \pm
0.02\,$(stat) and $\sigma_{\pi^-} = 2.29 \pm 0.02$ rapidity units.
Similar overall features have already been observed in central Au+Au
collisions at AGS~\cite{piAGS} and Pb+Pb reactions at
SPS~\cite{kpiSPS1}. This is reminiscent of the hydrodynamical
expansion model proposed by Landau~\cite{landau}. In the initial
state, colliding nuclei are highly Lorentz contracted along the beam
direction. Under assumptions of full stopping and isentropic expansion
after the initial compression phase (where thermal equilibrium is
reached), the hydrodynamical equations (using the equation of state of
a relativistic gas of massless particles) lead to $dN/dy$
distributions of Gaussian shape at freeze--out~\cite{milekhin}. The
original Landau model was simplified and used by Carruthers {\it et
  al.} for the description of particle production in p+p
collisions~\cite{minh,carut,cooper}. In their model, the width
$\sigma$ of the pion distribution is simply formulated:
\begin{equation}
\sigma^2 = \ln{\gamma_{beam}} = \ln{\left(\sqrt{s}/2m_p\right)}
  \label{eq:width}
\end{equation}
where $m_p$ is the proton mass. To extend this formula to heavy ion
collisions, one might naively replace $\sqrt{s}$ by \snn{}.
Figure~\ref{fig:landau}(a) shows the experimental $dN/dy(\pi)$ and
Carruthers Gaussians at \snn{200} (using Eq.~\ref{eq:width} with the
condition that the integrals of these Gaussians must be equal to the
full--space yields extracted from the data).
Surprisingly, a discrepancy of only $\sim$ 5\% between the model and
the data is observed ($\sigma_{\text{Carrut.}} = 2.16$). The inset in
Fig.~\ref{fig:landau} shows the ratio
$\sigma_{\text{data}}/\sigma_{\text{Carrut.}}$ as a function of
\snn{}. At all energies, the widths $\sigma_{\text{data}}$ result from
Gaussian fits to the pion distributions.  While the difference between
theory and measurements is of the order of 10\% at most from AGS to
RHIC energies, it is worth noting that the overall systematic may not
be trivial. The ratio at RHIC is $\sim$ 15\% higher than at SPS. It is
furthermore interesting to note that in central Au+Au collisions at
\snn{200}, the original baryons lose 2 rapidity units on average from
the initial value $y_b = 5.36$~\cite{peter}. Not only is the degree of
transparency significantly different between AGS and RHIC, but the
relative rapidity loss $\left<\delta y\right>/y_b$ is about half
lower~\cite{peter}.

On the basis of the Landau's original hydrodynamic,
Bjorken~\cite{bjorken} proposed a scenario in which yields of produced
particles would be boost--invariant within a region around
mid--rapidity. In that approach, reactions are described as highly
transparent leading to a vanishing net--baryon density around
mid-rapidity and particle production from pair creation from the color
field in the central zone. This would result in a flat distribution of
particle yields around $y=0$. As mentioned, collisions at RHIC are
neither fully stopped nor fully transparent, although a significant
degree of transparency is observed. Consequently the overall $dN/dy$
distribution of pions is expected to consist of the sum of the
particles produced in the boost invariant central zone and the
particles produced by the excited fragments. The fact that the
observed distributions are flatter at mid-rapidity and wider than
those predicted by the Landau--Carruthers model might point in this
direction.\\

In summary, we have measured transverse momentum spectra and inclusive
invariant yields of charged meson $\pi^\pm$ and $K^\pm$. The ratios of
strange to non--strange mesons $K/\pi$ are well reproduced by the
hadron gas statistical model~\cite{braun} that assumes strangeness
equilibration at mid--rapidity. The excess of $K^+$ over $K^-$ yields
at higher rapidities can be explained by the increasing
baryo--chemical potential $\mu_B$ with rapidity. The widths of the
pion rapidity distributions are in surprisingly good agreement with a
hydrodynamic model based on the Landau expansion picture.\\

This work was supported by the division of Nuclear Physics of the
Office of Science of the U.S. DOE, the Danish Natural Science Research
Council, the Research Council of Norway, the Polish State Com. for
Scientific Research and the Romanian Ministry of Research.

\bibliography{bibliography}

\end{document}